\documentclass{emulateapj}
\usepackage{apjfonts}

\newcommand{\src}{G15.9+0.2}

\catcode`\@=11
\newcommand{\gapprox}{\mathrel{\mathpalette\@versim>}}
\newcommand{\lapprox}{\mathrel{\mathpalette\@versim<}}
\newcommand{\propapprox}{\mathrel{\mathpalette\@versim\propto}}
\newcommand{\@versim}[2]
  {\lower3.1truept\vbox{\baselineskip0pt\lineskip0.5truept
\ialign{$\m@th#1\hfil##\hfil$\crcr#2\crcr\sim\crcr}}}
\catcode`\@=12

\shorttitle{NEW YOUNG GALACTIC SNR G15.9+0.2}
\shortauthors{REYNOLDS ET AL.}

\begin{document}

\title{A New Young Galactic Supernova Remnant Containing a Compact
Object:  G15.9+0.2}


\author{Stephen P. Reynolds,\altaffilmark{1}
Kazimierz J. Borkowski,\altaffilmark{1}
Una Hwang, \altaffilmark{2}
Ilana Harrus, \altaffilmark{2}
Robert Petre, \altaffilmark{2}
and Gloria Dubner \altaffilmark{3}}

\altaffiltext{1}{Department of Physics, North Carolina State University,
  Raleigh NC 27695-8202; stephen\_reynolds@ncsu.edu} 
\altaffiltext{2}{NASA/GSFC, Code 660, Greenbelt, MD 20771}
\altaffiltext{3}{Instituto de Astronom\'\i a y F\'\i sica del Espacio (IAFE), CC 67, 
Suc. 28, 1428 Buenos Aires, Argentina}

\begin{abstract}

We identify the radio-emitting shell-type supernova remnant G15.9+0.2
as a relatively young remnant containing an X-ray point source that
may be its associated neutron star.  The integrated spectrum of the
remnant shell obtained from our 30 ks exploratory Chandra observation
shows very strong lines that require elevated element abundances from
ejecta, in particular of sulfur.  A plane-shock model fit gives
a temperature $kT = 0.9 \ (0.8, 1.0)$ keV, an ionization timescale $n_et
= 6 \ (4, 9) \times 10^{10}$ cm$^{-3}$ s, and a sulfur abundance of
2.1 (1.7, 2.7) times solar (90\% confidence limits).  Two-component
models with one solar and one enriched component are also plausible,
but are not well constrained by the data. Various estimates give a
remnant age of order $10^3$ yr, which would make G15.9+0.2 among the
dozen or so youngest remnants in the Galaxy. The sparse point source
spectrum is consistent with either a steep $\Gamma \sim$ 4 power law
or a $kT \sim$ 0.4 keV blackbody.  The spectrum is absorbed by a H
column density $N_H \sim 4 \times 10^{22}$ cm$^{-2}$ similar to that
required for the remnant shell.  The implied 2--9.5 keV source
luminosity is about $10^{33}$ ergs s$^{-1}$ for an assumed distance of
8.5 kpc consistent with the high absorption column.  We suggest that
the point source is either a rotation-powered pulsar or a compact
central object (CCO).

\end{abstract}


\section{Introduction}
\label{intro}

Supernovae and supernova remnants (SNRs) power Galactic turbulence and
cosmic rays, drive Galactic chemical evolution, and illustrate the
recent history of star formation.  However, the Galaxy suffers a
well-known shortage of young remnants.  We are sure of only a few,
whereas typical supernova rates (e.g., van den Bergh \& Tammann 1991)
predict three or more SNe per century, or 60 expected SNRs younger
than 2000 yr. If we have not misunderstood Galactic supernova rates,
star formation rates, or SNR evolution, then we are missing many young
remnants.

Some of the missing young remnants may be concealed among known, but
poorly studied, small-diameter radio remnants.  We have begun a
program to identify young SNRs by conducting observations with the
{\sl Chandra} X-ray Observatory of several remnants with high radio
surface brightnesses in order to obtain spectra to estimate ages and
to search for evidence of ejecta.  The first results of that program
are presented in this paper.  The compact radio-bright remnant
G15.9+0.2 reveals enhanced elemental abundances and has an inferred
age of a few thousand years or less, making it one of the dozen or so
youngest SNRs in the Galaxy.  It is among the highest surface
brightness radio SNRs in the Galaxy \citep[see refs.~in][]{green06},
and has a fairly symmetrical radio shell structure
(Fig.~\ref{radioxim}), both of which properties are indications of
relative youth.  We have also made the unexpected discovery of a
central compact source in \src\ that is likely to be associated with
this remnant.  If this is the case, we can identify \src\ as the
remnant of a core-collapse supernova.

The distance to \src\ is poorly known, but there are several
indications of a large distance, including a high absorption column
density (verified by our spectral fits) and an estimate of 14 kpc from
the (untrustworthy) $\Sigma-D$ relation (Caswell et al.~1982,
corrected to a Galactocentric distance of 8.5 kpc).  In the absence of
any other information, we scale all results for a fiducial distance of
that to the Galactic Center, 8.5 kpc.  For this nominal assumed
distance, the mean remnant radius is 6.2 pc.

\section{Observations and Imaging}

We observed \src\ with {\sl Chandra} using the ACIS-S CCD camera (S3
chip) on 2005 May 23 (10 ks), May 25 (5 ks), and May 28 (15 ks).  We
checked the aspect correction, removed pixel randomization in energy,
corrected for charge transfer inefficiency, and filtered the light
curve to reject background flares.  We used calibration data from
CALDB version 3.1.0.  For spectral analysis, data were binned to a
minimum of 25 counts per bin to allow the use of $\chi^2$ statistics.
Our source extraction region contained 32,000 counts (19,700 after background
subtraction). Background was taken
from essentially the entire area of the S3 chip not occupied by the
source and not too near the chip edge.  The relatively high percentage
of background reflects Galactic Ridge emission, along with the
relatively low surface brightness of \src.

\begin{figure}
\epsscale{1.4}
\plotone{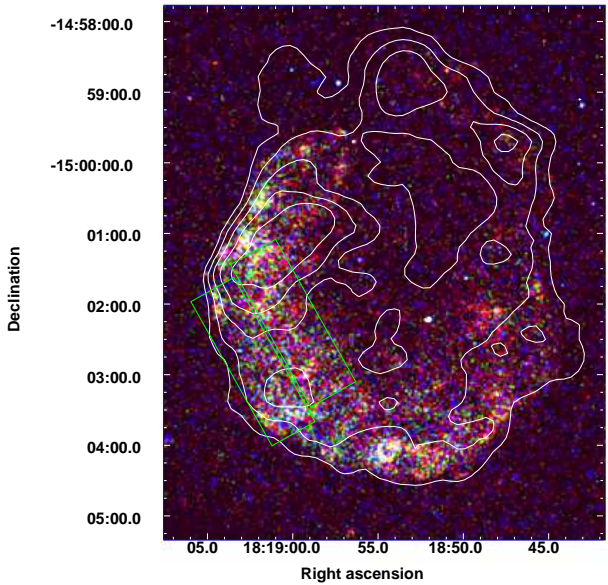}
\caption{{\sl Chandra} image of \src, with superposed radio contours.
  Red:  1 -- 2 keV.  Green:
2 -- 3 keV.  Blue:  3 -- 7 keV.  The image has been convolved with
a $2''$ FWHM Gaussian.  Contours:  1.4 GHz VLA image, resolution
$12'' \times 5''$. Inner and outer-shell regions 
described in the text are shown.  The compact central source is the
bright object slightly SW of the shell center, at about
$(18^{\rm h}18^{\rm m}52^{\rm s}, \ -15^\circ 02'14'')$.
\label{radioxim}}
\end{figure}

Figure~\ref{radioxim} shows the Chandra image of \src, with radio
contours. 
The shell is almost complete, with a very sharp outer edge. 
An unresolved source (hereafter CXOU
J181852.0-150213) can be clearly seen in the interior, though well
offset from the remnant center.
The radio image 
was produced after re-processing archival data obtained in 1993 at 1385
and 1465 MHz with the VLA in two hybrid configurations BnC and CnD
\citep{dubner96}.
In addition to the 1.4 GHz image shown in Fig.~\ref{radioxim}, we have
also processed VLA data acquired in the direction to G15.9+0.2 at
327.5 and at 5 GHz. No evidence for a radio counterpart for the
central point source, or for a radio flat-spectrum pulsar wind
nebula, is evident in any of these radio images.

\section{Spectral Analysis of SNR Shell}

\begin{figure}
\epsscale{0.5}
\centerline{
\includegraphics[angle=270,scale=0.32]{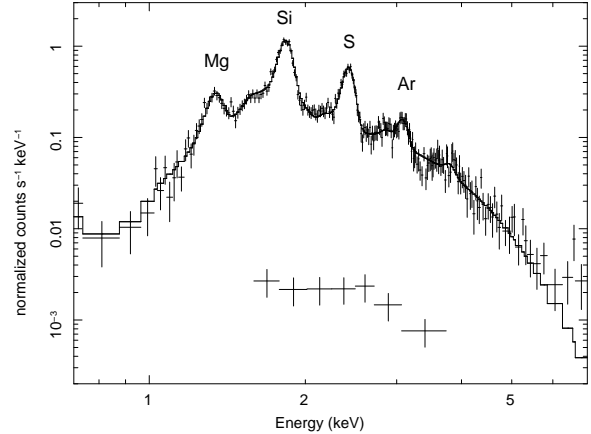}}
\caption{Integrated spectrum of \src, binned to a minimum of 25
counts per channel.  K$\alpha$ lines of several elements are labeled.
Model: plane shock with variable abundances ({\it vpshock}).  Lower
points: spectrum of central source, binned to 15 counts per channel.
\label{totspec}}
\end{figure}

The integrated remnant spectrum is shown in Fig.~\ref{totspec}.
Clearly apparent are strong K$\alpha$ emission features of Mg, Si, S,
Ar, and Ca, and the effects of high interstellar absorption.  The
integrated remnant spectrum cannot be adequately fitted with a model
for a planar shock with solar abundances \citep[XSPEC model {\it
pshock};][]{borkowski01}, but a model with variable abundances is much
more successful (see Fig.~\ref{totspec}; $\chi^2$ moved from 378 to
245 for 233 degrees of freedom).  The fitted column density of $N_H =
(3.9 \pm 0.2) \times 10^{22}$ cm$^{-2}$ is relatively insensitive to
the model. (Errors throughout are 90\% confidence intervals.)  The
electron temperature and ionization timescale are $kT_e = (0.9 \pm
0.1) $ keV and $n_e t = 5.4 (4,8) \times 10^{10}$ cm$^{-3}$ s.  This
model has a peculiar abundance pattern, requiring S and Ar at 2 -- 5
times solar, but solar Si.  Such an anomaly might arise if the
spectrum were more complex than assumed, for instance if there were
actually two thermal components.  We can obtain good fits
($\chi_\nu^2 = 1.1$) with two components, one with solar abundances and
the other with all abundances co-varied (in solar ratios).  Solar
abundances in both components are ruled out at a high level of
significance ($\Delta \chi^2 = 97$ with 235 degrees of freedom),
but the
amount of enhancement is strongly correlated with the emission measure
(the product of these quantities being roughly constant).  The fitted
absorption is consistent with the value given above for a single
component.  The solar and elevated-abundance components have
temperatures of 0.7 (0.6, 0.9) and 2.3 (1.9, 3) keV, respectively, and
ionization timescales of $7 (4, 12) \times 10^9$ and $6 (4, 8) \times
10^{10}$ cm$^{-3}$ s.  
The poorly constrained emission measure of the hotter component 
is an order of magnitude below that of the
cool component for the 90\%-confidence lower limit on the enhancement
(a factor of ten).  The rms electron density implied by the dominant
cool component $EM \equiv \int n_e n_H dV$ of $1.0 \times
10^{59}(D/8.5 \ {\rm kpc})^2$ cm$^{-3}$ is similar to that of the
one-component model, about 3.8 $(D/8.5 \ {\rm kpc})^{-1/2}$
cm$^{-3}$.

\begin{figure}
\epsscale{0.6}
\centerline{
\includegraphics[width=7truecm,height=5truecm]{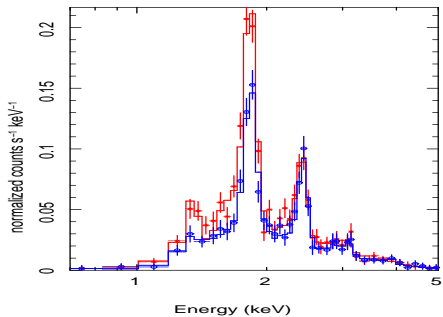}}
\caption{Spectra of outer (blue circles) and inner (red points) regions
in remnant SE.  Note clear spectral differences.
\label{2fits}}
\end{figure}

Supporting evidence for spatial variations in spectra comes from
separate examination of two regions in the SE of the shell: one
including the outer edge of emission, of width about $40''$ in the
radial direction, and a second one inside the first, also of width
$40''$ (Fig.~\ref{radioxim}).  Fits to these two regions using {\it
vpshock} (elements Mg, Si, S, Ar, and Ca allowed to vary in abundance)
show significant differences (Fig.~\ref{2fits}).  In particular, the
fitted ionization timescale for the outer region is $n_et = 2 \ (0.5,
8) \times 10^{11}$ cm$^{-3}$ s, six times larger than that of the
interior region.  While the errors on fitted parameters are
substantial, a fit to one spectrum when applied to the other increased
$\chi^2$ by 81 after renormalization (92 degrees of freedom; reduced
$\chi^2$ increased from 0.5 to 1.3). Furthermore, there is a
suggestion of higher overabundances in the inner region: S about three
times solar, Ar about six times, but with substantial errors.  While
these results do not constitute firm evidence for a distinct ejecta
component, they do suggest it, and indicate spectral complexity in the
remnant that is consistent with the relatively young age we infer
below.  A more complete understanding of the thermal shell emission in
\src\ will require a deeper observation, but this is not necessary for
our general conclusions.

\section{CXOU J181852.0-150213}

The CXOU J181852.0-150213 emission is consistent with a point source,
and contains about 120 total counts in our observation (after
background subtraction).  The spectrum is highly absorbed with an
$N_H$ value consistent with that of the shell.  We therefore froze
$N_H$ at the shell value of $3.9 \times 10^{22}$ cm$^{-2}$ for fitting
of the CXOU J181852.0-150213 spectrum.  We used Gehrels weighting of
the data, appropriate for our count rates \citep{gehrels86}, and
binned the data to at least 15 counts per bin, resulting in only 7
bins (we have no evidence of counts below 1 keV).  (Churazov weighting
[Churazov et al.~1996], an alternative method for low count-rate data,
gave similar results.)  A power-law fit gives a photon index $\Gamma
\sim 3.8\ (3.1, 4.3)$. 
A blackbody fit
with $kT = 0.4 \pm 0.1 $ keV is equally acceptable.  The inferred
source flux is relatively insensitive to spectral parameters: $S_x
\sim 1 \times 10^{-13}$ erg cm$^{-2}$ s$^{-1}$ (2 -- 9.5 keV) giving
an X-ray luminosity in that energy range of $L_x \sim 10^{33} (D/8.5 \
{\rm kpc})^2$ ergs s$^{-1}$.  The total luminosity of a spherical
blackbody with $kT \sim 0.4$ keV is $L_{bb} = 3.3 \times 10^{33} (R/10
\ {\rm km})^2$ erg s$^{-1}$.  This luminosity is comparable to that of
other point sources associated with SNRs, as we discuss below.  The
time resolution of CCD full-frame imaging observations is only 3.2 s,
so it is impossible to search for the fast periods characteristic of
normal pulsars.  Furthermore, our sparse data are inadequate for
period searches on longer timescales.

The source position (fit with WAVDETECT) is $18^{\rm h}18^{\rm m}
52\fs081 \pm 0\fs006,\ -15^\circ 02' 13\farcs91 \pm 0\farcs06$
(J2000).  The quoted ($1 \sigma$) errors reflect the statistical
accuracy of the X-ray measurement only. The statistical error in the
X-ray coordinate system registration is also 100 mas; we measured the
position of a bright star visible in X-rays and shifted X-ray
coordinates by a total of $79 \pm 102$ mas based on its USNO UCAC2
position \citep{zacharias04}. The combined $1 \sigma$ statistical
error is 150 mas. No other bright X-ray sources coincident with stars
with accurate position are present in the {\it Chandra} field of view,
so we cannot improve the coordinate registration accuracy and check
for a potential rotation of {\it Chandra} images (but rotation errors
are generally negligible).

We searched various catalogs for optical or IR counterparts within
$2''$ of our best-fit position. There is a star $1\farcs0$ away to the
S in the GLIMPSE survey using {\sl Spitzer}'s IRAC camera (Benjamin et
al.~2003), G015.8774+00.1966, and another fainter star $1\farcs9$ to
the N (G015.8781+00.1970). These two stars have been detected in the
IRAC channels 1 and 2 only (with effective wavelengths of 3.6 and 4.5
$\mu$m), and not at longer wavelengths. Their $m_{[3.6]}$ and
$m_{[4.5]}$ magnitudes are $13\fm0$, $13\fm0$ (G015.8774+00.1966), and
$13\fm7$, $13\fm8$ (G015.8781+00.1970). At {\sl Spitzer}'s spatial
resolution these two stars blend together, with the X-ray source
located between them.  These two stars have no counterparts in the
2MASS catalog of point sources, but their blend is faintly visible on
images in all (J, H, and K$_S$) 2MASS bands. Nothing is visible at the
X-ray source location on either optical sky survey images or in the
{\sl Spitzer} MIPSGAL Survey of the Inner Galactic Plane at 24 and 70
$\mu$m.

The absence of a bright IR/optical counterpart rules out a stellar
coronal origin for the detected X-ray source. The registration of the
GLIMPSE and UCAC2 coordinate systems is accurate, as demonstrated by the
small (200 mas) difference between GLIMPSE and UCAC2 coordinates of
the reference star used to align the X-ray image. This difference is
consistent with the internal GLIMPSE $1 \sigma$ error of 100 mas for
this star. The measurement error for G015.8774+00.1966 is comparable,
so the total $1 \sigma$ statistical error in the coordinate difference
measurement between this star and the compact source does not exceed
200 mas. Because systematic errors are not expected to exceed
statistical errors, the spatial offset between the two GLIMPSE stars
and the X-ray source makes them unlikely candidates for the IR
counterparts.  Their presence in all 2MASS band images suggests that
they are located much closer than G15.9+0.2. The high ($4 \times
10^{22}$ cm$^{-2}$) column density $N_H$ toward G15.9+0.2 implies a
very high ($A_V=22^{\rm m}$) optical extinction, and expected
absorption is high even in the IR \citep[{$6\fm2$, $3\fm9$, $2\fm5$,
$1\fm4$, and $1\fm1$ in J, H, K$_S$, [3.6], and [4.5] photometric IRAC
bands, respectively; we used an IR extinction curve
of}][]{indebetouw05}.  Such high extinction would cause reddening by
$5\fm1$ between the J and $[3.6]$ bands, making detection of the faint
18$^{\rm m}$ IR counterpart in the J band unlikely (the 2MASS point
source catalog sensitivity limit in this band is $15\fm8$ at 99\%\
completeness in regions of low stellar density). The two GLIMPSE stars
are seen in all 2MASS bands, however, implying less extreme extinction
(and hence a smaller distance than for G15.9+0.2). This and a lack of
spatial alignment with the X-ray source strongly suggests that they
are just two foreground stars projected against the SNR. An
identification of an IR counterpart to the central compact X-ray
source in G15.9+0.2 is not possible with either 2MASS or GLIMPSE data
because of high spatial confusion in the Galactic Plane and the
relatively low sensitivity of these surveys.  Deep near-IR
observations with much better spatial resolution are necessary to
identify the IR counterpart to the central compact X-ray source in
G15.9+0.2.

Explanations other than the neutron star interpretation for CXOU
J181852.0-150213 are unlikely, but cannot be excluded without
additional observations. At the observed 2--10 keV X-ray flux of $5
\times 10^{-14}$ ergs cm$^{-2}$ s$^{-1}$, most X-ray sources seen
within the Milky Way are extragalactic but with a nonnegligible
contribution from low-luminosity Galactic X-ray sources such as
cataclysmic variables \citep{hands04}. We expect only 0.01 sources
with equal or larger fluxes in the central region of G15.9+0.2
encompassing CXOU J181852.0-150213, based on the count rate of 40 such
sources per deg$^{-2}$ measured by \citet{hands04}.  The power-law fit
to the spectrum of CXOU J181852.0-150213 excludes at the 90\%
confidence level the typical value of $\Gamma \sim 2$ we would expect
for an AGN, though a deeper observation would be desirable to confirm
this conclusion.  An accurate measurement of the interstellar
absorption would be very helpful in establishing the true nature of
CXOU J181852.0-150213. Unfortunately, the current 90\%\ confidence
ranges for $N_H$ are too broad ($[1.8-6.8] \times 10^{22}$ cm$^{-2}$ and
$[3.8-13] \times 10^{22}$ cm$^{-2}$ for fits with the black body and the
power law, respectively) to allow us to distinguish between various
alternatives.


\section{Discussion and Conclusions}

We may attempt to estimate the age of \src\ in various ways.  Most
directly, we assume an average time evolution of the shock radius of
$R_s \propto t^m$, and take $R_s = 6.2 (D/8.5 \ {\rm kpc})$ pc, and a
shock velocity on the order of $u_8 \equiv v_s$/(1000 km s$^{-1}$)
based on measured gas temperatures above 0.7 keV.  (In the absence of
electron-ion temperature equilibration, the shock velocity would be
higher, and the inferred age lower.)  Assuming $m = 0.4$ as for Sedov
evolution, we then find $t = 2400/u_8(D/8.5 \ {\rm kpc})$ yr.  If the
remnant has not fully reached the Sedov stage, then $m > 0.4$ and this
age is an upper limit.  We can also use the Sedov relation $R_s = 1.15
(E/\rho_0)^{1/5} t^{2/5}$ and an assumed explosion energy $E =
10^{51}$ erg to estimate $t = 1400 (D/8.5 \ {\rm kpc})^{9/4}$ yr.
From the emission measure for the
single-component fit, we estimate an upstream density of $n_0 \sim
0.7(D/8.5 \ {\rm kpc})^{-1/2}$ cm$^{-3}$ assuming a factor of four
compression at the shock.  The inferred swept-up mass is then $4\pi
R_s^3\rho_0/3 = 22 (D/8.5 \ {\rm kpc})^{5/2} \ M_\odot$, which is
comparable to the expected ejected mass for a core-collapse explosion,
as required for the assumption of Sedov dynamics to be
self-consistent.  However, it is not much larger than that expected
ejecta mass, consistent with a young age.

The upstream density and swept-up mass inferred from the
solar-abundance component of our two-component fits are very similar
to those found from the single-component fit, since the
elevated-abundance component contributes little to the total emission
measure.  The age may also be estimated from the ionization age and
the density as $t_s = n_et/n_{\rm rms} = 540(D/8.5 \ {\rm kpc})^{1/2}$
yr.  Since the fitted values of the ionization age are less
reliable than, say, $EM$, it is not surprising that this estimate is
somewhat different. 

We consider the nature of the point source located inside \src.
The compact source
in \src\ has properties that are typical of compact point sources
associated with supernova remnants: a blackbody temperature of 0.4 keV
and a luminosity of order $10^{33}$ ergs s$^{-1}$ \citep{pavlov04}.
An association with the remnant is strengthened because the high
absorption of the point source's spectrum implies that it and the
remnant are at a comparable distance.  The blackbody temperature,
however, is too high for a normal cooling neutron star of any age
\citep{yakovlev04}, and the low luminosity would require that the
emission arise from only about 0.3\% of the surface.  This is a
common property of the CCO [central compact object] class; in
particular, thermal emission from small ``hot spots'' on the surface
of a neutron star was proposed by Pavlov et al.~(2000) to explain the
low X-ray luminosity of the CCO in Cas A. Compared to anomalous X-ray
pulsars, our point source's luminosity is one to two orders of
magnitude too low \citep{woods04}, and our data do not allow us to
verify either the typical blackbody plus power law spectrum or the
long periodicities typical of AXPs.  If we have underestimated the
distance by a factor of three to ten, the luminosity would fall in the
AXP class, but that would require \src\ to be at least 25 kpc away --
an unlikely possibility.  Magnetospheric emission from
a rotation-powered pulsar may also explain the
compact central source in \src.  The observed $L_x$ is typical of
such objects, as is a power-law spectrum (though the slope in this
case is somewhat steep).  However, if the compact source is an active
pulsar, it might be expected to produce a pulsar-wind nebula (PWN).
We see no evidence for extended emission around the source, 
though, as with other issues, a considerably deeper observation would
be desirable.  We consider a CCO or rotation-powered pulsar to be
viable possibilities for the point source in G15.9+0.2.

If the compact source is indeed associated with \src, a somewhat high
space velocity is required assuming that it was born near the
geometric center of the shell emission.  Its current position is
offset by about $35''$ from a visual estimate of the shell center,
which translates to a sky-plane velocity of about $700 (D/8.5 \ {\rm
kpc})(t/2000 \ {\rm yr})$ km s$^{-1}$, where $t$ is the remnant age.
While high, this velocity is not implausible, since as many as 15\% of
pulsars may be born with the implied 3-D spatial velocity required of
over 1000 km s$^{-1}$ (Arzoumanian, Chernoff, \& Cordes 2002); the CCO
in Pup A has a directly measured velocity of order 1000 km s$^{-1}$
\citep{winkler06, hui06}.

We conclude that \src\ is probably no more than a
few thousand years old, making it one of the 10 or 20 youngest
remnants in the Galaxy.  The tidy shell morphology 
is consistent with a young age, and our spectral analysis
indicates enhanced abundances which are likely to be ejecta emission
(although ambient abundances at the Galactocentric radius of 2.4 kpc
for \src\ may be somewhat higher than locally).  Ejecta emission alone
does not demand extreme youth since several SNRs in the Galaxy and the
LMC show detectable ejecta even at ages of order $10^4$ yr
\citep[e.g.,][]{hughes95, hendrick03}.  Taken as a whole with the low
emitting mass and relatively low ionization timescale we are confident
that \src\ is a young remnant.

The compact object in \src\ may be a candidate for addition to the
very short list of CCO's in supernova remnants (Pavlov et al.~[2004]
cite six CCO's).  Its age of 1000--3000 yr is comparable to that of
several of the CCOs, as is its X-ray luminosity.  The richness
of the X-ray spectrum of the shell remnant indicates that a deeper
X-ray observation could allow improved inferences on the remnant age
and abundances.  It would also clarify the nature of the central
source, whether a PWN-less rotation-powered pulsar, a CCO, or a
background AGN.

\acknowledgments

We thank the anonymous referee for a particularly
thorough and helpful review.  This work was supported by NASA through
Chandra GO Program grant GO5-6051A.


\end{document}